\documentclass[11pt,paper]{JHEP3}
\usepackage{epsfig}
\usepackage{latexsym}
\usepackage{amssymb}
\usepackage{amsmath}
\usepackage{booktabs}
\usepackage{graphicx}
\usepackage{dcolumn}
\usepackage{bm}

\newcommand{\be}{\begin{equation}}
\newcommand{\ee}{\end{equation}}
\newcommand{\ba}{\begin{eqnarray}}
\newcommand{\ea}{\end{eqnarray}}
\newcommand{\bi}{\begin{itemize}}
\newcommand{\ei}{\end{itemize}}

\newcommand{\nn}{\nonumber \\}

\newcommand{\ud}{\,\mathrm{d}}

\newcommand{\bsig}{\boldsymbol{\sigma}}

\newcommand{\<}{\langle}
\renewcommand{\>}{\rangle}
\newcommand{\eq}{Eq.~}

\newcommand{\la}{\label}

\newcommand{\ello}{{\bar\ell}}
\newcommand{\bam}{\bar{m}}
\newcommand{\txts}{\textstyle}

\newcommand{\psif}{\psi_{\rm f}}
\newcommand{\psii}{\psi_{\rm i}}
\newcommand{\bep}{\boldsymbol{\epsilon}}

\newcommand{\bB}{\boldsymbol{B}}
\newcommand{\bS}{\boldsymbol{S}}
\newcommand{\bP}{\boldsymbol{P}}
\newcommand{\bL}{\boldsymbol{L}}
\newcommand{\bA}{\boldsymbol{A}}
\newcommand{\bX}{\boldsymbol{X}}
\newcommand{\bY}{\boldsymbol{Y}}
\newcommand{\bJ}{\boldsymbol{J}}
\newcommand{\boe}{\boldsymbol{e}}
\newcommand{\bR}{\boldsymbol{R}}

\newcommand{\bn}{\boldsymbol{n}}
\newcommand{\bj}{\boldsymbol{j}}
\newcommand{\bp}{\boldsymbol{p}}

\newcommand{\br}{\boldsymbol{r}}
\newcommand{\bk}{\boldsymbol{k}}
\newcommand{\kg}{k}
\newcommand{\bkg}{\boldsymbol{k}_\gamma}
\newcommand{\bx}{\boldsymbol{x}}

\title{Photodisintegration of a Bound State on the Torus}

\author{Harvey B.\ Meyer \\
Institut f\"ur Kernphysik\\
Johannes Gutenberg-Universit\"at Mainz\\
D-55099 Mainz, Germany\\
\email{meyerh@kph.uni-mainz.de}}

\abstract{ In this article the cross-section for the
  photodisintegration of a bound state is expressed, order by order in
  the multipole expansion, in terms of matrix elements between states
  living on the three-dimensional torus. The motivation is to make the
  process amenable to Monte-Carlo simulations. The case of the
  deuteron is discussed.}

\begin{document}
\section{Introduction}

A non-perturbative quantum field theory such as quantum chromodynamics
(QCD) can presently be approached from first principles using
Monte-Carlo methods, provided the theory is formulated in a finite
spatial volume and in Euclidean time (see for
example~\cite{Gattringer:2010zz} for an introduction). The former
requirement is necessary in order to have a finite number of degrees
of freedom to handle on a computer, and the second allows one to
apply importance sampling methods. The spatial manifold is usually
chosen to be a torus, which has the advantage of preserving
translation invariance, at the cost of breaking the SO(3) rotational
symmetry down to the discrete symmetry group of the cube.

This setup is ideally suited to study the properties of the low-lying
spectrum of the theory. However it is not immediately clear how to
study scattering processes in this way, both because the scattering
states are discrete in finite volume (with energy gaps that are in
practice larger than the mass gap of the theory), and because the
correlation functions require an analytic continuation before they can
provide information on scattering phases and matrix elements involving
timelike momenta~\cite{Maiani:1990ca}. In a numerical approach
however, the analytic continuation represents an ill-posed problem,
see e.g.\ \cite{BerteroI}.

In a series of seminal
papers~\cite{Luscher:1986pf,Luscher:1990ux,Luscher:1991cf}, L\"uscher
demonstrated that this limitation could be overcome for elastic
scattering processes by relating the discrete two-particle spectrum on
the torus to the scattering phases of the two particles (see also the
older article~\cite{Huang:1957im}, where the problem is addressed
perturbatively in non-relativistic quantum mechanics).  This analytic
control over the two-particle spectrum allows one to extract other
dynamical properties from stationary observables, in particular
transition matrix elements from a one-particle
state~\cite{Lellouch:2000pv} or the QCD vacuum~\cite{Meyer:2011um} to
a two-particle scattering state. The derivation of the respective
formulas given in these articles involves considering the maximal
mixing between a one-particle state with a two-particle state on the
torus under the influence of a perturbing Hamiltonian. However there
are processes where a derivation along those lines is not possible (to
the best of the author's knowledge), and the technical motivation of the present
paper is to nonetheless derive a formula allowing one to extract a
scattering matrix element from stationary observables on the torus in
a different way.

The process we look at is the photodisintegration of a bound state
into a two-particle final state. Photoionization of an atom and
photodisintegration of light nuclei are classic reactions in atomic
and nuclear
physics~\cite{Friedrich,BlattWeisskopf,Arenhoevel:1991pa}, and one
reason we choose to study this type of reaction is because in these
physics contexts it can be described in non-relativistic quantum
mechanics, which is technically simpler. Indeed, except for the
absorption of the photon, the particle numbers are conserved.  We will
generalize the quantum mechanical model (\ref{eq:H1}) studied by
L\"uscher on the torus to include the perturbative interaction of the
particles with an electromagnetic field.  The main technical novelty
of this paper is the derivation of the relative normalization of a
partial wave contributing to a finite-volume state and the same
partial wave in infinite volume, \eq(\ref{eq:normLnormoo}). The main
application follows from \eq(\ref{eq:main}), which allows one in
principle to calculate the photodisintegration of an $s$-wave bound
state in the dipole approximation from numerical calculations on the
torus. An important lesson one learns from the derivation is that the
naive correspondence between the finite-volume states and the
infinite-volume states must be examined individually for each process.

Detmold and Savage investigated specifically the problem of
calculating electroweak deuteron photodisintegration in lattice QCD
several years ago~\cite{Detmold:2004qn}. They worked out a method in
detail to determine the low-energy constants of the pionless effective
field theory (EFT($\pi\!\!\!/$), \cite{Chen:1999tn}) relevant to
magnetic dipole transitions. With these constants in hand, the
photodisintegration process is predicted via the effective field
theory. By contrast, we provide a method to directly compute the
low-multipole matrix elements from lattice QCD, at the cost of having
more stringent requirements on the volume of the torus.

More broadly, the finite-volume methods to extract resonance
properties have attracted a lot of attention in recent years, and it
is appropriate to mention some of the work published recently.  Ways
to extend the formalism developed by L\"uscher to include the effects
of multiple open channels have been
proposed~\cite{Liu:2005kr,Lage:2009zv,Bernard:2010fp,Doring:2011vk}.
In~\cite{Luu:2011ep}, interpolating operators that are designed to
couple to a resonance of given quantum numbers are analyzed in detail.
At the same time there have been several recent numerical lattice QCD
calculations of the interactions among pions in the $\rho$
channel~\cite{Aoki:2007rd,Feng:2010es,Aoki:2011yj,Lang:2011mn}, and of
interactions in the two-baryon
sector~\cite{Yamazaki:2011nd,Beane:2011iw}.

The outline of this paper is as follows. In section (\ref{sec:theo}),
we review the basic relations between the spectrum of states on the
torus and the scattering phases and derive a relation between the
wavefunctions in finite and infinite volume.  Section (\ref{sec:EMoo})
is devoted to coupling a two-particle quantum mechanics model on the
torus perturbatively to photons, and the formula allowing one to
calculate the photodisintegration process order by order in the
multipole expansion is derived under stated assumptions.  In section
(\ref{sec:QCD}) the treatment is generalized so as to make contact
with potential applications in QCD. We conclude on the prospects of
such calculations and on future directions of interest.

\section{Relation between finite- and infinite-volume states\la{sec:theo}}

Consider the discrete energy levels of the physical system consisting
of two spinless particles of mass $m$ on the $L\times L\times L$
torus. We consider states in the $A_1$ and in the $T_1$ cubic
representation.  Under assumptions specified at the end of the
paragraph, the energy levels are related to their infinite-volume
scattering phase $\delta_{\ello}(k)$ via~\cite{Luscher:1986pf,Luscher:1990ux}
\be \la{eq:quantn}
\delta_{\ello}(k) + \phi(q) = n\pi, \qquad n\in\mathbb{Z},\qquad q\equiv \frac{kL}{2\pi},
\ee
where $\phi(q)$ is a known kinematic function\footnote{It is defined by $\tan\phi(q) =
-\frac{\pi^{{3}/{2}}q}{{\cal Z}_{00}(1;q^2)}$, where ${\cal
  Z}_{00}(1;q^2)$ is the analytic continuation in $s$ of ${\cal
  Z}_{00}(s;q^2) = \frac{1}{\sqrt{4\pi}} \sum_{\boldsymbol
  n\in\mathbb{Z}^3} \frac{1}{(\boldsymbol{n}^2-q^2)^s}$.
} defined and investigated in Ref.\ \cite{Luscher:1991cf}, Appendix A.
The lowest partial wave $\ello$ is 0 and 1 respectively in the $A_1$ and $T_1$ representations.
The `effective momentum' $k$ is defined from the energy of the state via the standard dispersion
relation, $E = \frac{k^2}{2\mu}$ in the non-relativistic case, $\mu$ being the 
reduced mass of the two particles. The relation (\ref{eq:quantn}) is strictly
exact for a central potential that vanishes identically beyond a radius $R$ (with 
$R<L/2$) and for angular momenta $l>\Lambda$,
\ba\la{eq:H1}
H &=& -\frac{1}{2\mu}\triangle + Q_\Lambda V(r).
\ea
For a wavefunction with spherical components $\psi_{lm}(r)$,
\be
\psi(\br) = \sum_{\ell=0}^\infty \sum_{m=-\ell}^\ell Y_{\ell m}(\theta,\phi) \psi_{\ell m}(r),
\ee
the operator $Q_\Lambda$ acts as follows,
\be
Q_\Lambda \psi(\br) = \sum_{\ell=0}^\Lambda 
\sum_{m=-\ell}^\ell Y_{\ell m}(\theta,\phi) \psi_{\ell m}(r).
\ee
The relation (\ref{eq:quantn}) is valid for $\Lambda$ strictly smaller
than the second partial wave contributing in a given irreducible
representation of the cubic group. Realistically, this means that it
is a good approximation as long as the higher partial waves are small,
which is guaranteed at sufficiently low energy. The relation
generalizes in various ways.  It remains correct up to terms
exponentially small in $L$ for a short-range potential (for example, a
Yukawa potential). Remarkably, it also remains valid up to
exponentially small terms in $L$ in the relativistic
case~\cite{Luscher:1990ux}, except that $k$ is then related to the
energy level via the relativistic dispersion relation
$E=2\sqrt{m^2+k^2}$. Finally, a generalization of (\ref{eq:quantn}) to
arbitrary values of $\Lambda$ exists~\cite{Luscher:1990ux}.

It is interesting to consider how the energy level changes under a
change in the box size. \eq(\ref{eq:quantn}) implies
\be \la{eq:dkdL}
\frac{L}{k} \frac{\partial k(L)}{\partial L} = 
-  \frac{q \phi'(q)}{F_{\ello}(k,L)}.
\ee
where 
\be
F_\ell(k,L) \equiv {k \frac{\partial \delta_\ell(k)}{\partial k} + q\phi'(q)}.
\ee
The quantity $F_\ell(k,L)$ that plays an important role in calculating
the decay of an unstable particle on the torus~\cite{Lellouch:2000pv,Meyer:2011um},
appears repeatedly in the equations below.  \eq(\ref{eq:dkdL})
provides a practical way of calculating it.

On the other hand, one may consider a perturbation to the potential that leads
to a change in the scattering phase $\Delta\delta_{\ello}(k)$, for a given
value of $k$. In a periodic box of fixed size, the allowed value of
$k$ must satisfy \eq(\ref{eq:quantn}) both before and after the
perturbation is added, for the same value of $n$. The difference of
the two relations results in
\be \la{eq:DkDd}
{\Delta \delta_{\ello} (k)} = - F_{\ello}(k,L) \frac{\Delta k}{k}  .
\ee
This relation will be used in the next section.

\subsection{Relation between the finite-volume and the infinite volume wavefunction}

In this section we derive a relation between the finite-volume
wavefunction $\Psi(\br)$ in a given irreducible cubic representation
${\cal R}$ and the wavefunction $\psi(\br)$ of the infinite-volume
scattering state, in the lowest partial wave $\ello$ that contributes to
${\cal R}$. The starting point is the Hamiltonian (\ref{eq:H1}) and we
choose $\Lambda$ such that \eq(\ref{eq:quantn}) holds.  Consider the
wavefunction $\psi_{\ell m}$ in the lowest partial wave contributing
to the given irreducible representation of the cubic group. In the
interaction region $r<R$, there is a unique solution to the radial
Schr\"odinger equation which remains finite at the origin and whose
normalization is required to be such that $r^{-\ell} \psi_{\ell
  m}=1$. This then determines the value and derivative of the radial
wave function at $r=R$. Since there are exactly two linearly
independent solutions to the radial Schr\"odinger equation, the
matching conditions at $r=R$ determine the two coefficients uniquely.
Therefore the radial wavefunction in the lowest relevant partial wave
is completely identical up to overall normalization to the
corresponding infinite volume radial wavefunction. In the following we
show how to calculate their relative normalization for a given
normalization of the full wavefunctions. 

We denote a finite-volume wavefunction by $\Psi(\br)$, and an
infinite-volume wavefunction by $\psi(\br)$.  We assume that the
former has unit norm,
\be
\int_{\Omega_L} \ud^3\br |\Psi(\br)|^2 = 1,
\ee
where $\Omega_L$ refers to one periodic cell of dimensions $L\times L\times L$.
In infinite volume, let the wavefunction take the form
\be
\psi(\br) = \sum_{\ell=0}^\infty \sum_{m=-\ell}^\ell Y_{\ell m}(\theta,\phi)\, \psi_{\ell m}(r).
\ee
Following the notation of~\cite{Luscher:1990ux}, we write 
\be
\psi_{\ell m}(r) = b_{\ell m} u_\ell(r;k),
\ee
with $u_\ell(r;k)$ a solution to the radial Schr\"odinger equation 
\be
\left(\frac{\ud^2}{\ud r^2}+\frac{2}{r}\frac{\ud}{\ud r} - \frac{l(l+1)}{r^2} + k^2 -2\mu V(r)
 \right)\psi_{\ell m}(r)=0
\ee
normalized according to
\be
\lim_{r\to0} r^{-\ell} u_\ell(r;k) = 1
\ee
(in other words, $b_{\ell m} = \lim_{r\to 0} r^{-\ell} \psi_{\ell m}(r)$).
The general solution for $r>R$ then reads
\be
u_{\ell}(r;k) = \alpha_\ell(k) j_\ell(kr) + \beta_\ell(k) n_\ell(kr)
\ee
for some coefficients $\alpha_\ell(k)$ and $\beta_\ell(k)$.
The scattering phase $\delta_\ell$ is related to them by 
\be
e^{2i\delta_\ell} = \frac{\alpha_\ell(k)+i\beta_\ell(k)}{\alpha_\ell(k)-i\beta_\ell(k)}.
\ee
Now for each $\ell$ we choose the values of the $b_{\ell m}$ such that 
\be\la{eq:blm1}
b_{\ell m} (\alpha_\ell(k)+i\beta_\ell(k)) = \delta_{m\bam} \sqrt{\frac{2\mu k }{\pi}} e^{i\delta_\ell}
\ee
for one particular value of $\bam\in\{-\ello,\dots,+\ello\}$.
The asymptotic $r\to\infty$ form of the radial wavefunction is then $\psi_{\ell \bam}(r) 
= \sqrt{\frac{2\mu k}{\pi}} \frac{1}{kr}\sin(kr-\ell\frac{\pi}{2}-\delta_\ell)$.
This implies that the states are `energy-normalized', i.e.\ two states
of energy $E$ and $E'$ in one particular partial wave $\ell$
normalized as in \eq(\ref{eq:blm1}) satisfy ($E=\frac{k^2}{2\mu}$, $E'=\frac{{k'}^2}{2\mu}$)
\be
\int \ud^3\br\; \psi(\br;\ell,E)^* \psi(\br;\ell',E') =
|b_{\ell\bam}|^2\delta_{\ell\ell'} \int_0^\infty \ud r\;r^2 \, u_\ell(\br;k)^* u_\ell(\br;k') 
= \delta_{\ell\ell'}\delta(E-E').
\ee

If the interaction potential is varied (while remaining short-range), $V\to V+\Delta V$,
the discrete energy levels in the box will change. On the other hand,
the phase shift also changes in infinite volume.  These two changes
have to be compatible with the L\"uscher relation (\ref{eq:quantn})
between finite-volume spectrum and phase shifts.

The change in a non-degenerate energy level is given by first-order perturbation 
theory of quantum mechanics\footnote{On the torus, the relevant potential is 
$V_L(\br) = \sum_{\bn\in \mathbb{Z}^3} V(|\br+\bn L|)$ and similarly for $\Delta V_L$.
We choose the angular momentum cutoff $\Lambda$ between the lowest and the next-to-lowest 
partial wave in a cubic irrep.},
\be\la{eq:I}
\Delta E = \int_{\Omega_L} \ud^3\br\; \Psi(\br)^* Q_\Lambda\Delta V_L(\br) \,\Psi(\br).
\ee
On the other hand, the change in the phase shift is given by 
the generalized Born formula (see e.g.\ \cite{LandauLif3}, parag.\ 133), 
which for our normalization of the wavefunction takes the form
\be\la{eq:II}
\Delta\delta_{\ello}
=-\pi \int_0^\infty r^2\ud r \, \Delta V(r) |\psi_{\ello \bam}(r)|^2.
\ee
The variation of $E$ and of $\delta_{\ello}$ must be related through \eq(\ref{eq:DkDd})
with ${\Delta k}= \frac{\ud k}{\ud E}\,\Delta E$
and $\frac{\ud k}{\ud E}$ given by the dispersion relation.
Combining (\ref{eq:I}, \ref{eq:II} and \ref{eq:DkDd}), one finds 
\be\la{eq:master}
\int_0^\infty r^2\ud r\, \Delta V(r) |\psi_{\ello \bam}(r)|^2 =
\frac{\ud k}{\ud E}\, \frac{F_{\ello}(k,L)}{\pi k} \int_{\Omega_L} \ud\br\; 
\Psi(\br)^* Q_\Lambda\Delta V_L(r) \Psi(\br).
\ee
This equation is valid for any potential $\Delta V$ of range $R<L/2$.

We will exploit in particular \eq(\ref{eq:master}) in the $T_1$ representation
for $\ello=1$ and $\Lambda = 2$. We can for instance choose a sufficiently short range
potential $\Delta V$, in which case one obtains the relation between the slopes 
at the origin of the two wavefunctions\footnote{
Since the lowest partial wave dominates in any case at small $r$, the angular momentum
cutoff is not needed in \eq(\ref{eq:master}) if $\Delta V$ is sufficiently short-range.}.
More generally we obtain in the $\ell=1$ channel the relation 
($\Psi_{\ell m}(r) \equiv \int \ud\Omega\; Y_{\ell m}(\theta,\phi)^*\; \Psi(\br)$)
\be\la{eq:normLnormoo}
|\psi_{1,\bam}(r)|^2 = \frac{\ud k}{\ud E}\cdot \frac{F_1(k,L)}{\pi k} 
\cdot |\Psi_{1,\bam}(r)|^2, \qquad\qquad (r<L/2).
\ee
To summarize, this equation confirms the statement that the finite-
and infinite-volume wavefunctions are simply related to each other.
The infinite volume, energy-normalized $\ell=1$ wavefunction is
proportional to the $\ell=1$ component of the finite-volume,
unit-normalized wavefunction $\Psi$ in the $T_1$ representation.  The
proportionality coefficient can be determined from purely
spectroscopic measurements on the torus. Finally, we will choose the
phase convention that $\psi_{10}(r)$ and $\Psi_{10}(r)$ are both real.

\subsection{Higher partial waves}

In the region where the potential vanishes, the absolute normalization of the 
higher partial waves contributing to the finite-volume wavefunction can be determined.

We consider again the $T_1$ representation for $\bar\ell=1$ and $\Lambda = 2$.
In the `outer' $R<r<L/2$ region, where $V(r)$ vanishes, the wavefunction
is the solution of the Helmholtz equation~\cite{Luscher:1990ux},
\be
\Psi(\br) = v_{1\bam}\; G_{1\bam}(\br,k^2) .
\ee
The radial function multiplying $Y_{1\bam}(\theta,\phi)$ is~\cite{Luscher:1990ux}
\be\la{eq:E1}
\Psi_{1\bam}(r) = -\frac{k^2}{4\pi} v_{1\bam} 
\left[ n_1(kr) + {\cal M}_{1\bam,1\bam}\, j_1(kr) \right]
\ee
On the other hand, we have seen that the radial wave function is proportional 
to the radial function in infinite volume.
Comparing \eq(\ref{eq:E1}) with \eq(\ref{eq:normLnormoo}), 
one recovers L\"uscher's quantization condition,
\be
\alpha_1 = {\cal M}_{1\bam,1\bam}(q) \beta_1
\ee
as well as the condition that determines the normalization factor
$v_{10}$.  Parametrizing generically $\alpha_\ell=\rho_\ell \cos\delta_\ell$,
$\beta_\ell=\rho_\ell \sin\delta_\ell$, one finds
\be
v_{1\bam} = -\sqrt{\frac{2\mu}{F_1(k,L)} \,\frac{\ud E}{\ud k}}\; 
\frac{4\pi}{k}\sin\delta_1.
\ee

For $\ell>\Lambda$ the partial wave of the state $\Psi(\br)$ is given by 
\ba\la{eq:PsiEllM}
\Psi_{\ell m}(r) 
&=& -\frac{k^2}{4\pi} v_{1\bam}\; {\cal M}_{1\bam,\ell m}\; j_\ell(kr)
=  \sqrt{\frac{2\mu}{F_1(k,L)} \,\frac{\ud E}{\ud k}}\, k\sin\delta_1\,
  {\cal M}_{1\bam,\ell m}\;  j_\ell(kr)
\ea
while the infinite-volume, energy-normalized wave function is given by
\be
\psi_{\ell m}(r) = \delta_{m\bam}\sqrt{\frac{2\mu k}{\pi}} \; j_\ell(kr)
\ee
since we neglect scattering in higher partial waves. 

\section{A simple two-particle system and radiation\la{sec:EMoo}}

We now couple the two-particle system discussed in the previous
section to the photon. Much of this section is textbook
material~\cite{Friedrich,BlattWeisskopf}, but serves as the
preparation for the situation on the torus.  Consider first, somewhat
more generally, a non-relativistic $N$-body system coupled to the
electromagnetic field,
\ba\la{eq:Hn}
H_{\rm kin} &=& \sum_{c=1}^N\frac{1}{2m_c}\left(\bp_c-e_c \bA(\br_c)\right)^2 
\\ &= & \frac{1}{2M} \left(\bP-{\txts\sum_c} e_c \bA_c\right)^2 
 + \frac{1}{4M}\sum_{c,d=1}^N m_c m_d
\left({\txts\frac{1}{m_c}}\left(\bp_c - e_c\bA_c\right) 
    - {\txts\frac{1}{m_d}}\left(\bp_d - e_d\bA_d\right)\right)^2,
~~~~
\ea
where $M = \sum_{c=1}^N m_c$, $\bP = \sum_{c=1}^N \bp_c$
and we abbreviate $\bA(\br_c)\equiv \bA_c$.
We will restrict ourselves to the case of two particles, whose properties
we index by $a$ and $b$.
Introducing as usual the center-of-mass and relative coordinates,
\be
\bR = \frac{m_a\br_a+m_b\br_b}{M},\qquad \br = \br_b-\br_a\qquad (M=m_a+m_b),
\ee
the Hamiltonian (\ref{eq:Hn}) can be written in the form
\be
H_{\rm kin}\ = \frac{1}{2M} \left(\bP- (e_a\bA_a + e_b\bA_b)\right)^2 
+ \frac{1}{2\mu}  \left(\bp - \mu 
({\txts\frac{e_b}{m_b}}\bA_b-{\txts\frac{e_a}{m_a}} \bA_a )\right)^2
\ee
where $\mu = \frac{m_am_b}{m_a+m_b}$ is the reduced mass of the two
particles.  We note that the non-vanishing commutation relations are
now $[P_i,R_j]=[p_i,r_j]=-i\delta_{ij}$.

We will be looking at one-photon processes and therefore only need the
terms linear in the vector potential.  We define
\ba
H_I = -\frac{1}{2M} \{\bP,e_a\bA_a + e_b\bA_b\},
\qquad\qquad 
h_I = -\frac{1}{2} 
\{\bp, ({\txts\frac{e_b}{m_b}}\bA_b - {\txts\frac{e_a}{m_a}}\bA_a)\}
\ea
where the curly braces represent anticommutators.  Introducing a
quantization box of size $L_q$ (which is unrelated to $L$ and will be
sent to infinity at the end), the gauge field is expanded in plane-wave
eigenmodes,
\be\la{eq:plane_wave_exp}
\bA(t,\br) = \sum_{\bk,\sigma} \sqrt{\frac{2\pi}{\omega_k L_q^{3}}}
\left(a_{\bk,\sigma} \bep_\sigma(\bk)\, e^{i(\bk\cdot\br-\omega t)}
+ {a_{\bk,\sigma}^\dagger} \bep_\sigma(\bk)^*\, e^{-i(\bk\cdot\br-\omega t)}\right),
\ee
with $[a_{\bk,\sigma},a_{\bk',\sigma'}] = \delta_{\bk\bk'}\delta_{\sigma\sigma'}$.
For long wavelengths compared to the size of the bound state,
we may expand the exponential. To calculate the anticommutator, 
we  use the relation 
$\{p_j,r_l\} = i\mu [H,r_jr_l] + \epsilon_{ilj} L_i$,
where  $\bL \equiv \br\times \bp$ is the orbital angular momentum operator,
and obtain
\be
\bep_{\sigma} \cdot \{\bp,\, e^{i\bk\cdot\br}\} = 2\bp\cdot\bep_{\sigma}
 - \mu [H,(\epsilon_\sigma\cdot\br)(\bk\cdot\br)] + i  \bL\cdot(\bk\times\bep_\sigma) +{\rm O}(k^2).
\ee
Thus 
\ba
h_I = -\frac{1}{2}\sum_{\bk,\sigma} \sqrt{\frac{2\pi}{\omega_kL_q^3}} 
&\Bigr\{& a_\sigma(\bk) e^{i(\bk\cdot\bR-\omega t)}
\Bigr(({\txts\frac{e_b}{m_b}-\frac{e_a}{m_a}}) 2i\mu [H_0,\br\cdot\bep_\sigma(\bk)]
\\ && 
 -\; \mu^2({\txts\frac{e_b}{m_b^2}+\frac{e_a}{m_a^2}}) 
\left[H_0,(\epsilon_\sigma(\bk)\cdot\br)(\bk\cdot\br)\right]
\nn && 
+\; i\mu ({\txts\frac{e_b}{m_b^2} + \frac{e_a}{m_a^2}}) \bL\cdot 
(\bk\times\bep_\sigma(\bk))+{\rm O}(k^2)\Bigr)+{\rm h.c.}\Bigr\}.
\nonumber
\ea
In the same way, one finds
\ba
H_I &=& -\frac{1}{M} \sum_{\bk,\sigma} \sqrt{\frac{2\pi}{\omega_kL_q^3}} \Bigr\{ a_\sigma(\bk) 
e^{i\bk\cdot\bR} (\bep_\sigma(\bk)\cdot \bP) \; 
\big(e_a e^{-i\frac{m_b}{M}\bk\cdot\br}+e_b e^{i\frac{m_a}{M}\bk\cdot\br}\big)+{\rm h.c.}\Bigr\}.\qquad~
\ea
In the center-of-mass frame, the matrix element of this term vanishes.

\subsection{Transition matrix element: dipole approximation}

We now want to study the cross-section for the photodissociation of a
bound state of particles $a$ and $b$. We assume for simplicity that
this bound state is a scalar (pure $s$-wave bound state).

Let us first assume that the wavelength of the photon is long compared
to the size $r_s$ of the bound state, $k_\gamma r_s\ll 1$. Then the leading
contribution to the matrix element of the interaction Hamiltonian in a
power expansion in $k_\gamma r_s$ is
\ba
\<\psif |  h_I | \psii; \bkg,\sigma\> &=& 
\Big\<\psif \Big| -i\mu{\txts\sqrt{\frac{2\pi}{\omega_\gamma L_q^3}}}  e^{i\bkg\cdot\bR}
({\txts\frac{e_b}{m_b}-\frac{e_a}{m_a}}) \,[H_0,\br\cdot\bep_\sigma(\bkg)]\, \Big|\psii \Big\>
+{\rm O}(k_\gamma)
\\ &=&
  -i\mu  {\txts\sqrt{\frac{2\pi}{\omega_\gamma L_q^3}}}
({\txts\frac{e_b}{m_b}-\frac{e_a}{m_a}}) \,
(E_{\rm f}-E_{\rm i}) \int \ud^3\br \; \psif(\br)^* (\bep_\sigma(\bkg)\cdot \br) \psii(\br)
+{\rm O}(k_\gamma).
\nonumber
\ea
This expression determines the cross-section in the dipole
approximation.  Using a definite angular momentum basis for the final
state, we see that only its $p$-wave component contributes.
According to Fermi's Golden Rule, the transition probability per unit
time is
\be
\frac{\ud P}{\ud t} = 2\pi |\<\psif |  h_I | \psii\>|^2 \rho(E_{\rm f}),
\ee
where $\rho(E)$ is the density of final states. For energy-normalized scattering states, 
this is just one. Calling 
\be
\int \ud^3\br \; \psif(\br)^* \,\br\, \psii(\br) \equiv  \br_{\rm fi},
\ee
the transition probability per unit time reads
\be
\frac{\ud P}{\ud t} = \frac{4\pi^2\mu^2}{L_q^3}({\txts\frac{e_b}{m_b}-\frac{e_a}{m_a}})^2
\omega_\gamma |\bep_\sigma(\bkg)\cdot\br_{\rm fi}|^2,
\ee
where we have used energy-conservation, $E_{\rm f} = E_{\rm i} +
\omega_\gamma$.  To obtain the cross-section, we must divide the transition rate 
by the photon flux. The latter is in the present case
one per volume $L_q^3$ times the speed of light, and the
cross-section for photodisintegration of the bound state is
\be\la{eq:sigma_dip}
\sigma_{\ell=1} =  {4\pi^2\mu^2} ({\txts\frac{e_b}{m_b}-\frac{e_a}{m_a}})^2
\omega_\gamma |\bep_\sigma(\bkg)\cdot\br_{\rm fi}|^2,
\ee
The key quantity to calculate is therefore the matrix element $\br_{\rm fi}$.

\subsection{Photodisintegration on the torus in the dipole approximation}

The treatment of the photodisintegration process laid out so far in
this section carries over to the torus with little change. One
difference is that the momenta accessible to the photons are discrete,
$\bkg=\frac{2\pi}{L}\bn$. The main question of interest here is whether
the infinite-volume matrix element $\br_{\rm fi}$ is accessible in the
finite-volume theory.

We work in the center-of-mass frame, $\bP=0$, and assume the two
particles to have equal masses. The initial bound state must therefore
be moving with a momentum $-\bkg$. In a non-relativistic treatment,
the wavefunction describing the relative motion inside the bound state
is independent of the total momentum, except for the torus boundary
conditions. The spatial boundary condition of the internal
wavefunction is either periodic or antiperiodic for equal-mass
particles~\cite{Rummukainen:1995vs}.  In the following paragraph, we
focus on the internal wavefunction of the system.

On the torus, both the initial and the final state internal
wavefunctions contain an infinite number of partial waves. The former
is in the $A_1$, the latter in the $T_1$ representation\footnote{As far as 
the initial state is concerned, if
  the periodic/antiperiodic boundary conditions break the cubic
  symmetry, the wavefunction belongs to an irreducible
  representation of the reduced symmetry group. This does not affect
  the discussion.}. However if the initial state is compact with a
radius $r_s$ a few times smaller than the box size, then the only
angular momentum component that is not exponentially suppressed in the
volume is the s-wave component\footnote{Indeed, the two linearly
  independent solutions to the free Schr\"odinger equation are
  $I_{\ell+\frac{1}{2}}$ and $K_{\ell+\frac{1}{2}}$.  The latter
  however diverges at the origin, and must therefore be excluded for
  $\ell>\Lambda$. That leaves $I_{\ell+\frac{1}{2}}$, which rises
  exponentially at large distances, say $e^{\kappa r}$.  But since all
  the partial waves must be of the same order at $r\approx L$ in order
  to satisfy the periodic boundary conditions, its coefficients must
  be of order $e^{-2\kappa L}$, since the s-wave radial wavefunction
  is of order $e^{-\kappa L}$ on the edge of the
  box.}~\cite{Luscher:1985dn,Koenig:2011ti}. Since the position
operator is a pure $\ell=1$ operator, the only partial wave that can
be reached is $\ell=1$. Furthermore, since the position-space
contributions to the matrix element are localized at $r<r_s$, the
matrix element $\boldsymbol{r}_{\rm fi}$ would be the same as in
infinite volume (up to exponential corrections), if the normalization
of the $p$-wave component of the final state were the same.  With the
normalization of states we have chosen, we find for the finite-volume
matrix element $\bR_{\rm fi}$, using \eq(\ref{eq:normLnormoo}),
\be\la{eq:Zfi}
|\epsilon_{\sigma}(\bkg)\cdot\br_{\rm  fi}|^2 
= \frac{\ud k}{\ud E} \cdot \frac{F_1(k,L)}{\pi k}\cdot  
|\epsilon_{\sigma}(\bkg)\cdot \bR_{\rm fi}|^2.
\ee
To repeat, the infinite-volume final state is meant to be a pure
$p$-wave, while the finite-volume state belongs to the $T_1$
representation.  \eq(\ref{eq:normLnormoo}) -- and therefore also
\eq(\ref{eq:Zfi}) -- assumes $m_a=m_b$.
Thus the cross-section for photodisintegration in infinite volume
(\ref{eq:sigma_dip}) can be expressed in terms of the finite-volume
matrix element $\bR_{\rm fi}$ up to exponentially suppressed
corrections.

\subsection{All-order multipole expansion}

If one does not expand in the momentum of the photon, the
photodisintegration cross-section in infinite-volume is proportional
to the square modulus of the matrix element
\ba\la{eq:hIme}
&& L_q^{3/2}\, \Big\<\psi_{\rm f}\Big| h_I \Big| \psi_i;\bkg,\sigma\Big\>\; e^{-i\bkg\cdot\bR} = 
\sqrt{\frac{\pi}{2\omega}} \cdot
\\
&& \qquad \qquad \int \ud^3\br\; \psi_{\rm f}(\br)^* 
\Big\{i\epsilon_\sigma(\bkg)\cdot \nabla, 
{\txts\frac{e_b}{m_b}}e^{i\frac{m_a}{M}\bkg\cdot \br} 
- {\txts\frac{e_a}{m_a}}e^{-i\frac{m_b}{M}\bkg\cdot \br}\Big \}
\psi_{\rm i}(\br).
\nonumber
\ea
The standard multipole expansion of a vector plane wave with a
wavevector in the $z$ direction reads~\cite{JDJackson}
\be\la{eq:vect_plane_wave_exp}
(\boe_x + i\sigma \boe_y) e^{i\kg z} = \sum_{\ell=1}^\infty c_\ell
\left[ j_\ell(\kg r) \bX_{\ell,\sigma} +\sigma \frac{1}{\kg} 
\nabla\times j_\ell(\kg r) \bX_{\ell,\sigma}\right],
\qquad \sigma=\pm1,
\ee
with $c_\ell = i^\ell \sqrt{4\pi(2\ell+1)}$.
For the reader's convenience we summarize in appendix the main
properties of vector spherical harmonics $\bY^M_{J\ell
  1}(\theta,\phi)$, of which the $\bX_{\ell m}$ are a special case.
Denoting $f_\ell^\sigma(\br) = \frac{1}{\sqrt{\ell(\ell+1)}}j_\ell(\kg r)
Y_{\ell, \sigma}(\theta,\phi)$, one can rewrite
\be\la{eq:ME2}
(\boe_x +i\sigma \boe_y) e^{i\kg z}=
\frac{1}{i\kg}\sum_{\ell=1}^\infty c_\ell \Big( -\sigma \nabla 
\Big( f^\sigma_\ell(\br) + \br\cdot\nabla f^\sigma_\ell(\br)\Big)
 -\sigma \kg^2 \br f^\sigma_\ell(\br)
+ \kg \br\times\nabla  f_\ell^\sigma(\br)
\Big)_{\br=z \boe_z}.
\ee
Now one can convince oneself that upon inserting the expansion
(\ref{eq:ME2}) into the matrix element (\ref{eq:hIme}), each term
labeled by the index $\ell$ connects the initial bound state to a
scattering state with orbital angular momentum given by
$\boldsymbol{L}^2|\psi_{\rm f}\> = \ell(\ell+1) |\psi_{\rm f}\>$.

On the torus, the final state is chosen in the $T_1$ representation.
Its wavefunction contains infinitely many partial waves. Since we have
related the normalization of the wave function in finite-volume to the
normalization in infinite volume (see \eq(\ref{eq:normLnormoo}) and
(\ref{eq:PsiEllM})), we can express the matrix element on the torus in
the following way,
\ba\la{eq:allorders}
\<\Psi^{T_1,m=\sigma}_{\rm f}| h_I | \Psi^{A_1}_{\rm i}; \bkg,\sigma\>
&=&  
\sqrt{\frac{\ud E}{\ud k} \frac{\pi k }{F_1(k,L)}} \Big(
\<\psi^{(\ell=1,m=\sigma)}_{\rm f} | h_I | \psi_{\rm i}; \bkg,\sigma\> 
\\ && 
+\sin\delta_1 \sum_{\ell=3}^\infty 
{\cal M}_{1\sigma,\ell \sigma}  \;
\<\psi^{(\ell,m=\sigma)}_{\rm f} | h_I | \psi_{\rm i}; \bkg,\sigma\>
 \Big),
\nonumber
\ea
Each term in this equation corresponds to a multipole amplitude.
Furthermore, every term in the multipole expansion of the interaction
Hamiltonian (\ref{eq:hIme}, \ref{eq:vect_plane_wave_exp}) contributes
exactly to one term in the series (\ref{eq:allorders}).  From a
practical point of view, the matrix elements calculable on the torus
thus correspond to a linear combination of infinite-volume matrix
elements to final states of definite angular momentum with calculable
coefficients. These coefficients are $L$ dependent, therefore by
measuring the finite-volume matrix element at several $L$ values and
fitting them to a truncated version of formula (\ref{eq:allorders}), one
can in principle determine multipole the matrix elements in several
low-lying partial waves, provided such a truncation in $\ell$ is
justified.

We note that if the scattering phases for $\ell\geq3$ are not
negligible, the coefficients of the series (\ref{eq:allorders})
change, but its structure remains the same, because in each partial
wave, the radial wavefunction is proportional to the corresponding
infinite-volume radial wavefunction.

\section{Potential applications in QCD\la{sec:QCD}}

In this section we investigate to what extent the results of section
\ref{sec:theo} and \ref{sec:EMoo} can be applied to the two-nucleon
system in QCD.

\subsection{Transition matrix element of a general electromagnetic current}

We now allow the charge and current density $(\rho,\bj)$ to have a general form.
The linear part of the interaction with the photon field then reads, in Coulomb gauge,
\be\la{eq:Hint} 
h_I+H_I \to H_{\rm int} = -\int \ud^3\bx\; \bj(\bx)\cdot\bA(\bx).  
\ee 
For $ \bj(\bx) =
\frac{1}{2}\sum_c \frac{e_c}{m_c} \big\{\bp_c,\delta(\bx-\br_c)\big\}
$ one recovers the model of section \ref{sec:EMoo}.  From that
preparatory discussion it is clear that the relevant matrix element
for the process of photodisintegration is
\be\la{eq:FVme}
\Big\<\Psi_{\rm f}\Big| 
\int \ud^3\bx\; \epsilon_\sigma(\bkg)\cdot\bj(\bx)\; e^{i\bkg\cdot\bx} 
\Big| \Psi_{\rm i}\Big\>.
\ee
The multipole expansion (\ref{eq:vect_plane_wave_exp}) of the
photon plane wave can just as well be applied to this more general
interaction.  The lowest order term in $k_\gamma$ comes from the gradient
term in \eq(\ref{eq:ME2}).  Generically, when the photon field is a
pure gauge, $\bA=\nabla G$, the interaction Hamiltonian gives
\be
-\int \ud^3\bx\; \bj(\bx)\cdot\nabla G(\bx) 
= -i \Big[H,\int \ud^3\bx \; \rho(\bx)\; G(\bx)\Big].
\ee
In the present case $G= f^\pm_\ell(\br) + \br\cdot\nabla
f^\pm_\ell(\br)$ is an eigenfunction of the operator $\bL^2$ with
eigenvalue $\ell(\ell+1)$.  At low momentum transfer, the dominant
term is $\ell=1$, and therefore the partial wave excited in $\Psi_{\rm
  f}$ is $\ell=1$. The relation between the finite-volume matrix
element (\ref{eq:FVme}) and the infinite-volume matrix element at low energies 
is therefore the same as in \eq(\ref{eq:Zfi}),
\ba\la{eq:main}
&& \Big|\Big\<\psi_{\rm f}^{(\ell=1,\sigma)}\Big| 
\int \ud^3\bx\; \epsilon_\sigma(\bkg)\cdot\bj(\bx)\; e^{i\bkg\cdot\bx} 
\Big| \psi_{\rm i}\Big\>\Big|^2 
\\ && = 
\frac{\ud k}{\ud E} \frac{F_1(k,L)}{\pi k}\cdot 
\Big|\Big\<\Psi^{T_1,\sigma}_{\rm f}\Big| 
\int_{\Omega_L} \ud^3\bx\; \epsilon_\sigma(\bkg)\cdot\bj(\bx)\; e^{i\bkg\cdot\bx} 
\Big| \Psi_{\rm i}\Big\>\Big|^2
+{\rm O}(k_\gamma).
\nonumber
\ea
We remind the reader that the normalization of the infinite-volume
scattering state in this equation is
$\<\psi^{(\ell)}(E)|\psi^{(\ell')}(E')\> =
\delta_{\ell\ell'}\;\delta(E-E')$.

\subsection{Magnetic transition near threshold}

We have so far assumed that the two particles are spinless and
considered the transition from an $s$-wave bound state to a $p$-wave
scattering state, which is the dominant one at low energies, because
$J=0$ to $J=0$ transitions are forbidden according to the selection
rules for one-photon processes~\cite{BlattWeisskopf}. However in the
presence of spins, a transition $\ell=0$ to $\ell=0$ is possible, and
in fact dominates at sufficiently low energies, as has been known for
a long time.  Fermi discovered in 1935 that the radiative neutron
capture cross-section by hydrogen is non-negligible near
threshold~\cite{Fermi1935}, and provided the correct explanation. The
process is essentially a transition from a ${}^{2S+1}L_J= {}^1S_0$
scattering state to a ${}^3S_1$ bound state, the deuteron. The spins
of the proton and neutron play a central role in this very low energy
regime. In particular, the amplitude is non-vanishing only because the
potential in the spin-triplet channel is different from the potential
in the spin-singlet channel. This leads to $s$-wave radial
wavefunctions of different energy that are not orthogonal to
each other.

The capture cross-section is related to the disintegration
cross-section by $\sigma_{\rm dis} = \frac{\bk^2}{\bkg^2}\sigma_{\rm
  cap}$, where $\bk$ is the relative momentum of the proton and
neutron, and $\bkg$ is the momentum of the
photon~\cite{BlattWeisskopf}.  Here we will consider the
disintegration process $d+\gamma\to p+n$, which can be investigated on
the torus along the same lines as above by including the spin degrees
of freedom. We still ignore for simplicity the tensor force, which
would lead in particular to a $d$-wave component in the deuteron. A
contribution of the form
\be
\bj_M(\bx) = \frac{ie}{2}\left(\frac{\mu_a}{m_a}\bsig_a\times [\bp_a,\delta(\bx-\br_a)]  
    + \frac{\mu_b}{m_b}\bsig_b\times [\bp_b,\delta(\bx-\br_b)]  \right)
\ee
to the electromagnetic current, with the corresponding interaction Hamiltonian
\be
-\int \ud\bx\; \bj_M(\bx)\cdot \bA(\bx) = 
-\frac{e}{2}\left( \frac{\mu_a}{m_a}\bsig_a\cdot \bB_a 
+ \frac{\mu_b}{m_b}\bsig_b\cdot \bB_b  \right),
\ee
induces (at lowest order in the multipole expansion) a magnetic dipole
transition. Here $\mu_a$ and $\mu_b$ correspond to the magnetic
moments of the proton and neutron in Bohr magnetons, e.g.\ 2.78 for the proton. Denoting
the two-nucleon spin states by $\chi(S,m_s)$ (for instance, $\chi(1,1)
= \chi_a^+ \chi_b^+$ and $\chi(0,0) = \frac{1}{\sqrt{2}}( \chi_a^+
\chi_b^- - \chi_a^- \chi_b^+)$) the initial state with $m=1$ is of the
form $\underline{\psi}_{\rm i} = \psi_{\rm i}(r) \chi(1,1)$ and the
final state $\underline{\psi}_{\rm f} = \psi_{\rm
  f}(r)\chi(0,0)$. Decomposing the current into an isovector
($\bsig_n-\bsig_p$) and an isosinglet ($\bsig_n+\bsig_p$) part, one
easily checks that only the former makes a contribution to the
${}^3S_1\to {}^1S_0$ transition\footnote{Yet twisted boundary
  conditions cannot straightforwardly be used to calculate this
  cross-section at very low momenta, for the reasons given
  in~\cite{Sachrajda:2004mi}.}.

As in the case of the ${}^3S_1\to {}^0P_1$ transition driven by the
electric dipole operator, the ${}^3S_1\to {}^1S_0$ matrix element on
the torus can be related to the infinite-volume matrix element,
\ba
&& \Big|\Big\<\psi_{\rm f}^{(\ell=0)}\Big|
{\int}\ud^3\bx\; \epsilon_\sigma(\bkg)\cdot\bj(\bx)\; e^{i\bkg\cdot\bx} 
\Big| \psi_{\rm i}\Big\>\Big|^2
\\ && \qquad = \frac{\ud k}{\ud E} \frac{F_0(k,L)}{\pi k}\cdot 
\Big|\Big\<\Psi_{\rm f}^{A_1^+}\Big|
{\int}_{\Omega_L}\ud^3\bx\; \epsilon_\sigma(\bkg)\cdot\bj(\bx)\; e^{i\bkg\cdot\bx}  
\Big| \Psi_{\rm i}\Big\>\Big|^2
\nonumber.
\ea

\subsection{Probing the nucleon-nucleon potential}

We return to relation (\ref{eq:I}), which relates the energy shift due
to a change in the interparticle potential to the wave function for a
non-degenerate state. If we choose again the angular momentum cutoff
$\Lambda$ between the lowest and next-to-lowest partial wave allowed
by the cubic symmetry and choose the perturbing potential to have a range smaller than $L/2$, 
the relation is equivalent to, up to exponentially small corrections,
\be\la{eq:phi_ell}
\Delta E = \frac{\ud E}{\ud k}\; \frac{\pi k}{F_\ell(k,L)}\; 
\int_0^\infty \ud r \;  r^2 \Delta V(r) \,\psi_{\ell m}(r)^2,
\ee
where $\ell$ is the lowest partial wave contributing to the given
cubic representation ($\ell=0$ and 1 respectively in the $A_1$ and $T_1$ representations),
and $m\in\{-\ell,\dots+\ell\}$.
A determination of $\Delta E$ thus gives access to information on the infinite-volume
 wavefunction.
For instance, choosing the one-parameter family of Yukawa potentials,
\be
\Delta V(r) = g^2\; \frac{e^{-\mu r}}{4\pi r},
\ee
the Laplace transform of $r\,\psi_{\ell m}(r)^2$ is obtained by
varying $\mu$.  An inversion of the Laplace transform allows one to
evaluate the unperturbed potential via
\be
2\mu V(r) = \frac{\phi_{\ell}''(r)}{\phi_{\ell}(r)} 
        + k^2- \frac{\ell(\ell+1)}{r^2},\qquad 
\phi_\ell(r) \equiv r\,\psi_{\ell m}(r).
\ee
We note that the absolute normalization of the wavefunction is not
required to determine $V(r)$. In the case of two nucleons, let us
consider then a channel with a definite total spin $S$ and definite
orbital angular momentum $\ell$. We disregard here the effects of the
tensor force which mixes several orbital angular momenta into a state
of definite total angular momentum $J$, leaving this issue to future
investigation.  It may be unrealistic to invert the Laplace transform
numerically, but the wavefunctions obtained by solving
phenomenological nuclear potentials can be inserted into the
right-hand side of \eq(\ref{eq:phi_ell}) and compared to the left-hand
side calculated in lattice QCD Monte-Carlo simulations. In this way,
\eq(\ref{eq:phi_ell}) provides a consistency check between lattice QCD
and phenomenological approaches to the two-nucleon system. We expect
this basic idea to remain valid when the local potential $V(r)$ is
replaced by a non-local one.

There is no simple way of implementing an angular momentum cutoff on
the perturbing potential. In general the energy shift
(\ref{eq:phi_ell}) then corresponds to summing over $\ell$ the term
appearing on the right-hand side.  In order to have a simple
interpretation, the method must be applied in the regime where the
effective momentum $k$ is small in comparison with the range of both
the internucleon potential and the perturbing potential $\Delta
V(r)$. In this way the contribution of higher partial waves is
kinematically suppressed.

How can one implement the method at the quark level?  The quarks can
be given a Yukawa coupling to a massive, color-neutral scalar boson.
The nucleon mass itself will then shift by order $g^2$, and this
effect must be subtracted in order to consider the change in binding
energy of the two-nucleon state in \eq(\ref{eq:phi_ell}).  An
attractive feature of this method to probe the internucleon potential
is that only spectroscopic `measurements' are involved\footnote{This
  is to be contrasted with a recently proposed procedure to determine
  off-shell nucleon potentials that depend on the interpolating
  operator, see~\cite{Aoki:2009ji}.}.  However, very accurate
numerical data is required.

\section{Conclusion\la{sec:concl}}

We have derived a relation, \eq(\ref{eq:main}), which provides a
recipe to calculate numerically the amplitude for the
photodisintegration of a scalar bound state. The result is based for
one on the relation between the lowest partial wave contributing to
the finite-volume wavefunction and the infinite-volume wavefunction in
the same partial wave; they are identical~\cite{Luscher:1990ux} up to
overall normalization, and the latter is given by
\eq(\ref{eq:normLnormoo}). And secondly, the result relies on the
multipole expansion of the transition amplitude.

In order to comment on the prospects of exploiting relation
(\ref{eq:main}) for the photodisintegration of the deuteron in lattice
QCD, we recall the assumptions made. An important one is that the
initial bound state is well contained in the box. This assumption
ensures for instance that the matrix element of the dipole operator
between the initial bound state and the final scattering state is
entirely due to the $p$-wave component of the latter, and that this
contribution is not affected by the `cutoff' in the radial variable at
$r=L/2$. Secondly, we assumed that only the $p$-wave phase shift is
significant, which is only justified at low energies. This assumption
can be relaxed thanks to the general analysis performed
in~\cite{Luscher:1990ux}. Thirdly, the box size must be large enough
for the smallest momenta of the photon, i.e.\ the momentum transfer
between the initial and the final bound state, to be in the relevant
low-energy region.

The requirements on the physical volume of the torus are thus very
strong, since the deuteron is a very weakly bound state, $E_{\rm
  bind}\approx 2.2$MeV and the energy range where the dipole
approximation works well is below 100MeV. A box size of 10fm or more
is required. Approaching the problem from heavier quark masses, where
the scattering phases have more `natural' sizes, could be possible if
the deuteron remains bound, as suggested by recent lattice
calculations~\cite{Yamazaki:2011nd,Beane:2011iw} (some earlier
effective field theory estimates~\cite{Beane:2002vs,Epelbaum:2002gb}
had suggested that the deuteron unbinds quite soon as the light quark
masses are increased from their physical value).  Finally, apart from
the kinematic requirements, obtaining an acceptable signal-to-noise
ratio in the two-nucleon channel is a tremendous challenge in lattice
QCD, see for
instance~\cite{Beane:2009py,Yamazaki:2011nd,Beane:2011iw}.  As an
exploratory calculation, it may be interesting to apply the methods
proposed here in (quenched) lattice QCD at heavy quark masses and very
large volumes as in~\cite{Yamazaki:2011nd}. Perhaps the process can
also be investigated in effective theories along the lines
of~\cite{Borasoy:2006qn,Montway:2011qt}. Whether there is a choice of
boundary conditions that alleviates the strong requirements on the box
size deserves further investigation. In any case, we hope that the
techniques presented in this paper will be useful to address further
scattering processes.

Studying the disintegration of nuclear bound states through operators
other than the vector current is also of great interest. As already
emphasized in~\cite{Detmold:2004qn} in the lattice QCD context, the
axial current is a particularly important case, since it is relevant
to the disintegration of the deuteron induced by neutrinos (see for
instance~\cite{Ying:1989as}).  This reaction was instrumental in
determining the flavor of the neutrinos arriving from the Sun in the
SNO experiment~\cite{Ahmad:2002jz}.

\acknowledgments{I thank J.W.\ Negele, A.\ Rusetsky, A.\ Schwenk and C.\ Urbach 
for interesting discussions. I also thank M.\ Savage and S.\ Beane for comments
on the first version of this article.
This work was supported by the \emph{Center for Computational Sciences in Mainz}.}

\bibliographystyle{JHEP}
\bibliography{/home/meyerh/CTPHOPPER/ctphopper-home/BIBLIO/viscobib.bib}

\appendix 

\section{Vector spherical harmonics}

For the reader's convenience we briefly describe the properties of the
vector spherical harmonics and how the multipole expansion
\eq(\ref{eq:vect_plane_wave_exp}) comes about, following the treatment
of~\cite{BlattWeisskopf}.  The vector spherical harmonics
$\bY^M_{J\ell 1}(\theta,\phi)$ are vector functions of the
$(\theta,\phi)$ angles which are simultaneous eigenfunctions of the
operators $J_z$ and $J^2$,
\be
J_z= L_z+S_z,\qquad \bL \equiv -i \br\times \nabla, \qquad 
S_i  = \boe_i\times (.)
\ee 
and 
\be
J^2=J_x^2+J_y^2+J_z^2.
\ee
The eigenvalues are respectively $M$ and $J(J+1)$.
In addition, the $\bY_{J\ell 1}^M$ have parity $(-1)^\ell$.
Since $\bL$ and $\bS$ commute, they are constructed by the standard 
addition law of angular momentum,
\be
\bY_{J\ell 1}^M(\theta,\phi) = \sum_{m=-\ell}^\ell \sum_{m'=-1}^1
C_{\ell1}(J,M;m,m') \,Y_{\ell m}(\theta,\phi) \bep_{m'}\,,
\ee
where 
\be
\bep_1 = -\frac{1}{\sqrt{2}}(\boe_x+i\boe_y),\qquad 
\bep_0 = \boe_z,\qquad 
\bep_{-1} = \frac{1}{\sqrt{2}}(\boe_x-i\boe_y).
\ee
The vectors $\bX_{\ell m}$ appearing in \eq(\ref{eq:vect_plane_wave_exp}) 
correspond to the special case
\be
\bX_{\ell m}(\theta,\phi) \equiv \bY_{\ell\ell 1}^m(\theta,\phi) 
= \frac{1}{\sqrt{\ell(\ell+1)}} \bL Y_{\ell m}(\theta,\phi).
\ee
The vector spherical harmonics are orthonormal,
\be
\int \ud\Omega\; \bY^M_{J\ell1}(\theta,\phi)^*\; 
\bY^{M'}_{J'\ell'1}(\theta,\phi) = \delta_{JJ'} \delta_{MM'} \delta_{\ell \ell'}.
\ee
Since the vector spherical harmonics form a complete set, an arbitrary vector 
field can be expanded as 
\be\la{eq:gen_exp}
\bA(\br) = \sum_{\ell=0}^\infty \sum_{m=-\ell}^\ell
\Big[f_{\ell m}(r) \bY^m_{\ell\ell 1} + g_{\ell m}(r) \bY^m_{\ell,\ell+1,1}
+h_{\ell m}(r) \bY^m_{\ell,\ell-1,1} \Big]
\ee
The radial functions $f_{\ell m},g_{\ell m},h_{\ell m}$ are obtained from the 
integrals
\ba\la{eq:flm}
f_{\ell m}(r) &=& \int \ud\Omega\; 
\bY^m_{\ell\ell1}(\Omega)^* 
\cdot\bA(\br),
\\ \la{eq:glm}
g_{\ell m}(r) &=& \int \ud\Omega\; \bY^m_{\ell,\ell+1, 1}(\Omega)^* \cdot \bA(\br),
\\ \la{eq:hlm}
h_{\ell m}(r) &=& \int \ud\Omega\; \bY^m_{\ell,\ell-1,1}(\Omega)^* \cdot \bA(\br).
\ea

The commutation relation $[\nabla\times,\bJ]=0$ implies that the 
curl of an expression with some definite value of $J$ and $M$ must be a linear 
combination of vector spherical harmonics with the same $J$ and $M$.
In particular, for an arbitrary radial function $f(r)$, we have
\be
\nabla\times (f(r)\bX_{\ell m}) = f_+(r)\bY^m_{\ell,\ell+1,1}
 - f_-(r)\bY^m_{\ell,\ell-1,1},
\ee
where $f_\pm(r)$ are related to $f(r)$. Thus the expansion
(\ref{eq:vect_plane_wave_exp}) of the plane wave is of the form
(\ref{eq:gen_exp}). Since only $m=\pm1$ appears on the right-hand side
of \eq(\ref{eq:vect_plane_wave_exp}), each term is individually
orthogonal to $\boe_z$. Finally, the specific radial coefficients
$f,g,h$ of \eq(\ref{eq:plane_wave_exp}) can be obtained from the
integrals (\ref{eq:flm}--\ref{eq:hlm}).

\end{document}